\documentclass[twocolumn,showpacs,superscriptaddress,amsmath,amssymb]{revtex4}
\usepackage{graphicx}
\usepackage{dcolumn}
\usepackage{bm}

\def\bd{
\begin{document}} \def\ed{\end{document}}
\def\bmp{\begin{minipage}} \def\emp{\end{minipage}}
\def\bcc{\begin{center}} \def\ecc{\end{center}}     \def\npg{\newpage}
\def\beq{\begin{equation}} \def\eeq{\end{equation}} \def\hph{\hphantom}
\def\be{\begin{equation}} \def\ee{\end{equation}} \def\r#1{$^{[#1]}$}
\def\n{\noindent} \def\ni{\noindent} \def\pa{\parindent}
\def\hs{\hskip} \def\vs{\vskip} \def\hf{\hfill} \def\ej{\vfill\eject}
\def\cl{\centerline} \def\ob{\obeylines}  \def\ls{\leftskip}
\def\underbar#1{$\setbox0=\hbox{#1} \dp0=1.5pt \mathsurround=0pt
   \underline{\box0}$}   \def\ub{\underbar}    \def\ul{\underline}
\def\f{\left} \def\g{\right} \def\e{{\rm e}} \def\o{\over} \def\d{{\rm d}}
\def\vf{\varphi} \def\pl{\partial} \def\cov{{\rm cov}} \def\ch{{\rm ch}}
\def\la{\langle} \def\ra{\rangle} \def\EE{e$^+$e$^-$} \def\pt{p_{\rm t}}
\def\bitz{\begin{itemize}} \def\eitz{\end{itemize}}
\def\btbl{\begin{tabular}} \def\etbl{\end{tabular}}
\def\btbb{\begin{tabbing}} \def\etbb{\end{tabbing}}
\def\beqar{\begin{eqnarray}} \def\eeqar{\end{eqnarray}}
\def\\{\hfill\break} \def\dit{\item{-}} \def\i{\item}
\def\bbb{\begin{thebibliography}{9} \itemsep=-1mm}
\def\ebb{\end{thebibliography}} \def\bb{\bibitem}
\def\bpic{\begin{picture}(260,240)} \def\epic{\end{picture}}
\def\akgt{\cl{\bf ACKNOWLEDGMENTS}}
\def\fgn{\noindent{\bf\large\bf figure captions}}
\def\lan{\langle}
\def\ran{\rangle}
\def\p{\pi}
\def\ifmath#1{\relax\ifmmode #1\else $#1$\fi}%
\def\rc{\ifmath{{\mathrm{c}}}}
\def\cut{\ifmath{{\mathrm{cut}}}}
\def\rF{\ifmath{{\mathrm{F}}}}
\def\rK{\ifmath{{\mathrm{K}}}}
\def\rp{\ifmath{{\mathrm{p}}}}
\def\rt{\ifmath{{\mathrm{t}}}}
\def\LAB{\ifmath{{\mathrm{LAB}}}}
\def\cut{\ifmath{{\mathrm{cut}}}}
\def\beq{\begin{equation}}
\def\eeq{\end{equation}}

\newcommand{\cinst}[2]{$^{\mathrm{#1}}$~#2\par}
\newcommand{\crefi}[1]{$^{\mathrm{#1}}$}
\newcommand{\crefii}[2]{$^{\mathrm{#1,#2}}$}
\newcommand{\crefiii}[3]{$^{\mathrm{#1,#2,#3}}$}
\newcommand{\HRule}{\rule{0.5\linewidth}{0.5mm}}

\bd

\title{\boldmath Spin-Parity Analysis of $p\bar{p}$ Mass Threshold Structure in $J/\psi$ and $\psi^\prime$ Radiative Decays}


\author{\small
M.~Ablikim$^{1}$, M.~N.~Achasov$^{5}$, D.~Alberto$^{41}$,
D.J.~Ambrose$^{38}$, F.~F.~An$^{1}$, Q.~An$^{39}$, Z.~H.~An$^{1}$,
J.~Z.~Bai$^{1}$, R.~B.~F.~Baldini Ferroli$^{17}$, Y.~Ban$^{25}$
% ...
\\
\vspace{0.2cm}
(BESIII Collaboration)\\
\vspace{0.2cm} {\it
$^{1}$ Institute of High Energy Physics, Beijing 100049, P. R. China\\
% ...
$^{5}$ G.I. Budker Institute of Nuclear Physics SB RAS (BINP), Novosibirsk 630090, Russia\\
% ...
$^{17}$ INFN Laboratori Nazionali di Frascati , Frascati, Italy\\
% ...
$^{25}$ Peking University, Beijing 100871, P. R. China\\
% ...
$^{38}$ University of Rochester, Rochester, New York 14627, USA\\
$^{39}$ University of Science and Technology of China, Hefei 230026, P. R. China\\
% ...
$^{41}$ University of Turin and INFN, Turin, Italy\\
% ...
}}

\vspace{0.4cm}

\date{\today}

\begin{abstract}

A partial wave analysis of the $p \bar{p}$ mass-threshold
enhancement in the reaction $J/\psi\rightarrow\gamma p\bar{p}$ is
used to determine: its $J^{PC}$ quantum numbers to be $0^{-+}$; its
peak mass to be below threshold at $M=1832^{+19}_{-5}~{\rm
(stat.)}^{+18}_{-17}~{\rm (syst.)}\pm19~{\rm (model)} ~{\rm
MeV}/c^2$; and its total width to be $\Gamma<76~{\rm MeV}/c^2$ at
the $90\%$ C.L. The product branching ratio is measured to be
$B(J/\psi\rightarrow\gamma X(p\bar{p}))B(X(p\bar{p})\rightarrow
p\bar{p})=(9.0^{+0.4}_{-1.1}~{\rm (stat.)}^{+1.5}_{-5.0}~{\rm
 (syst.)}\pm2.3~{\rm (model)})\times10^{-5}$.
 A similar analysis performed on $\psi^\prime\rightarrow\gamma p\bar{p}$ decays shows,
 for the first time, the presence of a corresponding enhancement with a production rate relative to that for $J/\psi$ decays of
 $R=(5.08^{+0.71}_{-0.45}~{\rm (stat.)}^{+0.67}_{-3.58}~{\rm
(syst.)}\pm0.12~{\rm (model)})\%$.

\end{abstract}


\maketitle

An anomalously strong $p\bar{p}$ mass threshold enhancement was
first observed by the BESII experiment in the radiative decay
process $J/\psi\rightarrow\gamma p \bar{p}$~\cite{ppb_jixb} and was
recently confirmed by the BESIII and CLEO-c~\cite{ppb_bes3-CLEO-c}
experiments. Curiously, no apparent corresponding structures were
seen in near-threshold $p\bar{p}$ cross section measurements, in
$B$-meson decays~\cite{Bdecay},  in radiative $\psi^\prime$ or
$\Upsilon\rightarrow\gamma p\bar{p}$ decays~\cite{psip_upslon}, or
in $J/\psi\rightarrow\omega p \bar{p}$ decays~\cite{omegappb}. These
non-observations disfavor the mass-threshold enhancement attribution
to the effects of $p\bar{p}$ final state interactions
(FSI)~\cite{{julich,theory1,one-pion-exchange}}.

A number of theoretical speculations have been proposed to interpret
the nature of this structure
~\cite{julich,theory1,one-pion-exchange,theory,baryonium}. Among
them, one intriguing suggestion is that it is due to a $p\bar{p}$
bound state, sometimes called baryonium~\cite{baryonium}, an object
with a long history and the subject of many experimental
searches~\cite{eexp1}. The observation of the $p\bar{p}$ mass
threshold enhancement also stimulated an experimental analysis of
$J/\psi\rightarrow\gamma\pi^+\pi^-\eta^\prime$ decays, in which a
$\pi^+\pi^-\eta^\prime$ resonance, the $X(1835)$, was first observed
by the BESII experiment~\cite{x1835} and  recently confirmed with
high statistical significance by the BESIII
experiment~\cite{x1835_bes3}. Whether or not the $p\bar{p}$ mass
threshold enhancement and the $X(1835)$ are related to the same
source still needs further study; among these, spin-parity
determinations and precise measurements of the masses, widths and
branching ratios are especially important.

In this letter, we report the first partial wave analysis (PWA) of
the $p\bar{p}$ mass threshold structure produced via the decays of
$J/\psi\rightarrow\gamma p \bar{p}$ and
$\psi^\prime\rightarrow\gamma p\bar{p}$.  Data samples containing
$(225.2\pm2.8)\times 10^{6}$ $J/\psi$ events  and $(106\pm4) \times
10^{6}$ $\psi^\prime$ events~\cite{jpsi_psip_num}
 accumulated in the Beijing Spectrometer
(BESIII)~\cite{bes3} located at the Beijing Electron-Positron
Collider (BEPCII)~\cite{bes2} are used.

The cylindrical core of the BESIII detector consists of a
helium-gas-based drift chamber~(MDC), a plastic scintillator
Time-of-Flight system~(TOF), and a CsI(Tl) Electromagnetic
Calorimeter~(EMC), all enclosed in a superconducting solenoidal
magnet that provides a 1.0-T magnetic field.  The solenoid is
supported by an octagonal  flux-return yoke with resistive plate
counter muon identifier modules (MU) interleaved with steel plates.
The solid angle for the charged particle and photon acceptance  is
$93\%$ of $4\pi$, and the charged particle momentum and photon
energy resolutions at 1~GeV are $0.5\%$ and $2.5\%$, respectively.
The time resolution of TOF is 80~ps in the barrel and 110 ps in the
endcaps, and the $dE/dx$ resolution is $6\%$.

Charged-particle tracks in the polar angle range $|\cos\theta|<0.93$
are reconstructed from hits in the MDC. The TOF and $dE/dx$
information are combined to form particle identification confidence
levels for the $\pi$, $K$ and $p$ hypotheses; the particle type with
the highest confidence level is assigned to each track.
 Photon candidates are required to have an energy deposit of at least $25~{\rm MeV}$  in
 the barrel EMC ($|\cos\theta|<0.8$) and $50~{\rm MeV}$ in
 the endcap EMCs  ($0.86<|\cos\theta|<0.92$), and be
isolated from antiprotons by more than $30^\circ$.

Candidate $J/\psi\rightarrow\gamma p\bar{p}$ events are required to
have at least one photon and two charged tracks identified as a
proton and an antiproton. Requirements of $|U_{miss}| < 0.05 ~{\rm
GeV}$, where $U_{miss} = (E_{miss} - |P_{miss}|)$,
 and $P^{2}_{t\gamma} < 0.0005 ~({\rm GeV/}c)^2$, where
$P^{2}_{t\gamma} = 4|P_{miss}|^2 \sin^{2}\theta_{\gamma}/2$, are
imposed to suppress backgrounds from multi-photon events. Here
$E_{miss}$ and $P_{miss}$ are, respectively, the missing energy and
momentum of all charged particles, and  $\theta_{\gamma}$ is the
angle between the missing momentum and the photon direction. A
four-constraint (4C) energy-momentum conservation kinematic fit is
 performed to the $\gamma p\bar{p}$ hypothesis.  For
 events with more than one photon candidates, the combination with the
 minimum $\chi^2$ is used; $\chi^2<20$ is also required.
Since there are differences in detection efficiency  between data
and Monte Carlo (MC) simulated low-momentum tracks, we reject events
containing any tracks with momentum below $0.3~{\rm GeV/}c$.

The $p\bar{p}$
 mass spectrum for events that satisfy all of the above-listed criteria
  is shown in Fig.~\ref{Fig1}(a). There is a clear
signal of $\eta_c$, a broad enhancement around $M_{p\bar{p}}\sim
2.1~{\rm GeV/}c^2$, and a prominent and narrow low-mass peak at the
$p\bar{p}$ mass threshold, consistent with that reported by
BESII~\cite{ppb_jixb} and BESIII~\cite{ppb_bes3-CLEO-c}. The Dalitz
plot for selected events is shown in Fig.~\ref{Fig1}(b).

\begin{figure}
\includegraphics[width=3.4in]{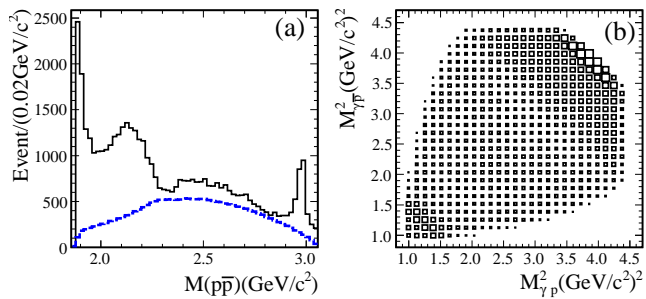}
\caption{\label{Fig1}  The $p\bar{p}$ invariant mass spectrum for
the selected $J/\psi\rightarrow\gamma p\bar{p}$ candidate events.
(a) The $p\bar{p}$ invariant mass spectrum; the open histogram is
data and the dashed line is from  $J/\psi\rightarrow\gamma p\bar{p}$
phase-space MC events(with arbitrary normalization). (b) An
$M^2(\gamma p)$~(horizontal) $versus$ $M^2(\gamma
\bar{p})$~(vertical) Dalitz plot for the selected events.}
\end{figure}

Potential background processes are studied with an inclusive MC
sample of $2\times 10^{8}$ $J/\psi$ events generated according to
the Lund model~\cite{lund}.
None of the background sources produces
an enhancement at the $p\bar{p}$ mass threshold region. The dominant
background is from $J/\psi\rightarrow\pi^{0} p\bar{p}$ events, with
asymmetric $\pi^{0}\rightarrow\gamma\gamma$ decays where one of the
photons has most of the $\pi^{0}$ energy. An exclusive MC sample,
generated according to the PWA results of $J/\psi\rightarrow\pi^{0}
p\bar{p}$ at BESII~\cite{jpsi_ppbpi0}, indicates that the level of
this background in the selected data sample with
$M_{p\bar{p}}<2.2~{\rm GeV}/c^2$ is $3.7\%$ of the total. The
$J/\psi\rightarrow\pi^{0} p\bar{p}$ decay channel is also studied
with data, and there is no evidence of a $p\bar{p}$ mass threshold
enhancement, which provides further evidence that the enhancement
observed in $J/\psi$ decays is not from background.

A PWA of the events with $M_{p\bar{p}}<2.2~{\rm GeV}/c^2$  is
performed to focus on determining the parameters of the $p\bar{p}$
mass threshold structure, which we denote as $X(p\bar{p})$. The
maximum likelihood method applied in the fit uses a likelihood
function that is constructed from $\gamma p\bar{p}$ signal
amplitudes described by the relativistic covariant tensor amplitude
method~\cite{ampt} and MC efficiencies.  The background contribution
from the $\pi^0 p\bar{p}$ process is removed by subtracting the
log-likelihood values of background events from that of data, since
the log-likelihood value of data  is the sum of the log-likelihood
values of signal and background events~\cite{lnl_bkg}. Here, the
background events are estimated by the MC sample of
$J/\psi\rightarrow\pi^0 p\bar{p}$ decays described above. We include
the  effect of FSI in the PWA fit using the Julich
formulation~\cite{julich}.

Four components, the $X(p\bar{p})$, $f_2(1910)$, $f_0(2100)$ and
$0^{++}$ phase space (PS) are included in the PWA fit. The
intermediate resonances are described by  Breit-Wigner (BW)
propagators, and the parameters of the $f_2(1910)$ and $f_0(2100)$
are fixed at PDG values.  In the optimal PWA fit, the $X(p\bar{p})$
is assigned to be a $0^{-+}$ state. The statistical significance of
the $X(p\bar{p})$ component of the fit is much larger than
 $30\sigma$;
those for the other components are larger than $5\sigma$, where the
statistical significance is determined from the changes of
likelihood value and degrees of freedom in the PWA fits with and
without the signal hypotheses. The mass, width and product branching
ratio (BR) of the $X(p\bar{p})$ are measured to be:
$M=1832^{+19}_{-5} ~{\rm MeV}/c^2$, $\Gamma=13\pm39~{\rm MeV}/c^2$
and $B(J/\psi\rightarrow\gamma X)B(X\rightarrow
p\bar{p})=(9.0^{+0.4}_{-1.1})\times10^{-5}$, respectively, where the
errors are statistical only. Figure ~\ref{Fig2} shows comparisons of
the mass and angular distributions between the data and the PWA fit
projections.  For the spin-parity determination of the
$X(p\bar{p})$, the $0^{-+}$ assignment fit is better than  that for
$0^{++}$ or other $J^{PC}$ assignments with statistical
significances that are larger than $6.8\sigma$.

\begin{figure}
\includegraphics[width=3.0in,width=3.4in]{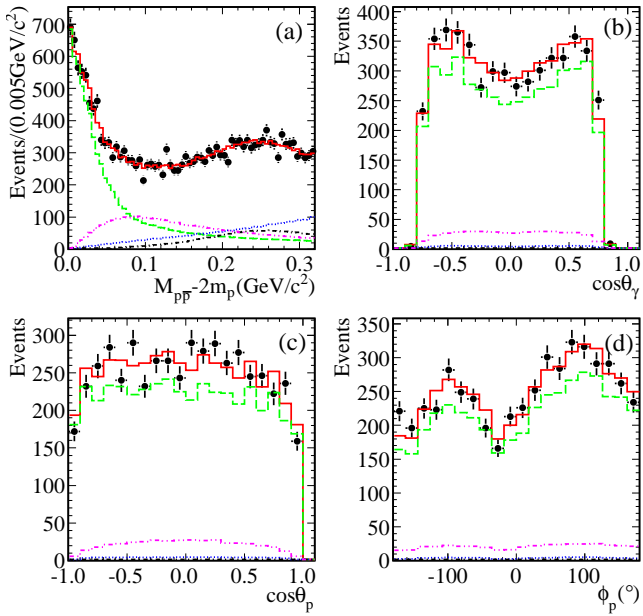}
\caption{\label{Fig2} Comparisons between data and PWA fit
projection: (a) the $p\bar{p}$ invariant mass; (b)-(d) the polar
angle  $\theta_\gamma$ of the radiative photon in the $J/\psi$
center of mass system, the polar angle  $\theta_p$ and the azimuthal
angle  $\phi_p$ of the proton in the $p\bar{p}$ center of mass
system with $M_{p\bar{p}}-2m_{p}<50~{\rm MeV}/c^2$, respectively.
Here, the black dots with error bars are data, the solid histograms
show the PWA total projection, and the dashed , dotted , dash-dotted
and dash-dot-dotted lines show
 the contributions of the $X(p\bar{p})$, $0^{++}$ phase
space, $f_0(2100)$ and $f_2(1910)$, respectively.}
\end{figure}

Variations of the fit included replacing the $f_0(2100)$ with the
$f_2(2150)$, the $f_2(1910)$ with the $f_2(1950)$, and replacing
both components simultaneously; changing the $J^{PC}$ of the PS
contribution, as well as consideration of the parameter
uncertainties of the $f_0(2100)$ and $f_2(1910)$, were performed,
and it is found the changes of the log-likelihood values and the
parameters of the $X(p\bar{p})$ are quite small, except that when
replacing $0^{++}$ PS with $0^{-+}$ PS the event number of the
$X(p\bar{p})$ decreases by $52\%$. We also tried fits that include
other possible resonances  listed in the PDG table~\cite{PDG2}
[$\eta_2(1870)$, $f_2(2010)$, $f_2(1950)$, $f_2(2150)$,
$f_{J}(2220)$, $\eta(2225)$, $f_2(2300)$, $f_2(2340)$ etc.] as well
as $X(2120)$ and $X(2370)$~\cite{x1835_bes3}, and  different
$J^{PC}$ PS contributions. The statistical significances of these
additional resonances are lower than $3\sigma$. All of the parameter
changes that are found in these alternative fits are considered as
sources of systematic uncertainties.

For systematic errors on the mass and width of the $X(p\bar{p})$,
 in addition to those discussed above, we include uncertainties from different fit ranges of
 $M_{p\bar{p}}<2.15~{\rm GeV}/c^2$ and $M_{p\bar{p}}<2.25~{\rm GeV}/c^2$,
different parameterizations for the BW formula, as well as different
background levels. For the systematic errors of the BR measurement,
there are additional uncertainties from the efficiencies of charged
track detection, photon detection and particle identification,
kinematic fit and the total number of $J/\psi$ events. The total
systematic errors on the mass and width of the $X(p\bar{p})$ are
$^{+18}_{-17}~{\rm MeV}/c^2$ and $^{+10}_{-13}~{\rm MeV}/c^2$,
respectively, and the corresponding relative systematic error on the
product BR is $^{+17}_{-56}\%$.

Various FSI models~\cite{julich,theory1,one-pion-exchange} have been
proposed to interpret the $p\bar{p}$ mass threshold enhancement.
Among them, a BW function times a one-pion-exchange FSI
factor~\cite{one-pion-exchange} can also describe the data well. For
this case, the mass and width of the $X(p\bar{p})$ shift by $19~{\rm
MeV}/c^2$ and $4~{\rm MeV}/c^2$, respectively, while the relative
change in the product BR is $25\%$. These errors are considered as
second (model) systematic errors due to the uncertainty of the model
dependence.

The $\psi^\prime\rightarrow\gamma p\bar{p}$ decay channel is also
studied using event selection criteria similar to those used in the
$J/\psi\rightarrow\gamma p\bar{p}$ study. The $p\bar{p}$ mass
spectrum of the surviving events is shown in Fig.~\ref{Fig3}(a).
Besides the well known $\eta_c$ and $\chi_{cJ}$ peaks, there is also
a $p\bar{p}$ mass threshold excess relative to PS. However, here the
line shape of the mass spectrum in the threshold region appears to
be less pronounced than that in $J/\psi$ decays. Potential
background processes were extensively studied with an inclusive MC
sample of $1\times10^8$ $\psi^\prime$ events and a data sample of
the selected $\psi^\prime\rightarrow\pi^0 p\bar{p}$ events, and
these indicate that the $p\bar{p}$ mass threshold structure is not
from any background source. An exclusive MC sample, generated
according to preliminary PWA results of $\psi^\prime\rightarrow\pi^0
p\bar{p}$ decays with BESIII data~\cite{jpsi_ppbpi0_bes3}, is
applied to the background estimation, and the background level from
this source in the selected data sample with $M_{p\bar{p}}<2.2~{\rm
GeV}/c^2$ is determined to be $3.4\%$.

\begin{figure}
\includegraphics[width=3.4in]{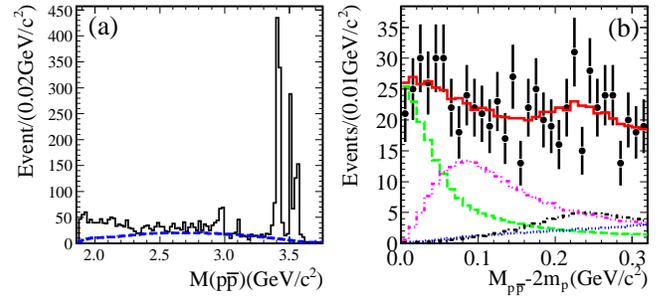}
\caption{\label{Fig3} (a) The $p\bar{p}$ invariant mass spectrum for
the selected $\psi^\prime\rightarrow\gamma p\bar{p}$ candidate
events; the open histogram is data and the dashed line is from a
$\psi^\prime\rightarrow\gamma p\bar{p}$ phase-space MC events(with
arbitrary normalization). (b)  Comparisons between data and PWA fit
projection for $p\bar{p}$ mass spectrum, the representations of the
error bars and histograms are same as those in Fig.~\ref{Fig2}. }
\end{figure}

A  PWA  on  $\psi^\prime\rightarrow\gamma p\bar{p}$ which is similar
to that applied for $J/\psi\rightarrow\gamma p\bar{p}$ decays was
performed to check the contribution of $X(p\bar{p})$ in
$\psi^\prime$ decays and   to measure the production ratio between
$J/\psi$ and $\psi^\prime$ radiative decays,
$R=B(\psi^\prime\rightarrow\gamma
X(p\bar{p}))/B(J/\psi\rightarrow\gamma X(p\bar{p}))$. Due to limited
statistics of $\psi^\prime$ events, in the PWA, the mass and width
of $X(p\bar{p})$ as well as its $J^{PC}$ were fixed to the results
obtained from $J/\psi$ decays.
 Figure ~\ref{Fig3}(b) shows
comparisons between data and MC projections for the $p\bar{p}$ mass
spectrum. As in $J/\psi$ decays, replacing the $f_0(2100)$  with the
$f_2(2150)$ and the $f_2(1910)$ with the $f_2(1950)$ yields no
significant change in fit quality. The determined product BR and $R$
value are $B(\psi^\prime\rightarrow\gamma X) \times B(X\rightarrow
p\bar{p})=(4.57\pm0.36)\times10^{-6}$ and
$R=(5.08^{+0.71}_{-0.45})\%$, respectively.

With the  consideration of systematic uncertainties  similar to
those in $J/\psi$ decays, and the uncertainty of the total number of
$\psi^\prime$ events, the total relative systematic error on the BR
is $(^{+27}_{-89}~{\rm (syst.)}\pm28~{\rm (model)})\%$ , and
systematic error on  R values is $(^{+0.67}_{-3.58}~{\rm
(syst.)}\pm0.12~{\rm (model)})\%$. Similar to all cases  studied in
$J/\psi$ analysis, the statistical significance of the $X(p\bar{p})$
is larger than $6.9\sigma$ in $\psi^\prime$ decays.

The PWA fits are also performed without the correction for FSI
effect. The corresponding log-likelihood value worsen  by 25.6 than
those with FSI effect included.
 The mass, width and product BR of the $X(p\bar{p})$ are
$M = 1861$ $\pm 1 {\rm (stat.)} $ $^{+13}_{-4} {\rm (syst.)}$ ${\rm
MeV}/c^2$,
 $\Gamma=1\pm 6 {\rm (stat.)}$ $^{+18}_{-1} {\rm (syst.)} ~{\rm MeV}/c^2$
 (a total width of $\Gamma<32 ~{\rm MeV}/c^2$ at the $90\%$ C.L),
  $B(J/\psi\rightarrow\gamma X(1860))$ $B(X(1860)\rightarrow p\bar{p})=(8.6^{+0.3}_{-0.2}~{\rm (stat.)}^{+2.4}_{-3.5}~{\rm
 (syst.)})\times10^{-5}$ and $B(\psi^\prime\rightarrow\gamma X(1860))$ $B(X(1860)\rightarrow p\bar{p})=(4.15\pm0.39~{\rm (stat.)}^{+2.51}_{-1.71}~{\rm
 (syst.)})\times10^{-6}$, respectively. The corresponding
 $R$ value is
 $(4.80^{+0.46}_{-0.48}~{\rm (stat.)}^{+2.24}_{-1.29}~{\rm (syst.)})\%$.

In summary, the PWA of $J/\psi\rightarrow\gamma p\bar{p}$ and
$\psi^\prime\rightarrow\gamma p\bar{p}$ decays are performed. In
$J/\psi$ radiative decays, the near-threshold enhancement
$X(p\bar{p})$ in the $p\bar{p}$ invariant mass is determined to be a
$0^{-+}$ state. With the inclusion of Julich-FSI effects, the mass,
width and product BR for the $X(p\bar{p})$ are measured to be:
$M=1832^{+19}_{-5}~{\rm (stat.)}^{+18}_{-17}~{\rm
 (syst.)}\pm19~{\rm
 (model)}~{\rm MeV}/c^2$, $\Gamma=13\pm39~{\rm
(stat.)}^{+10}_{-13}~{\rm
 (syst.)}\pm4~{\rm
 (model)}~{\rm MeV}/c^2$ (a total width of
$\Gamma<76~{\rm MeV}/c^2$ at the $90\%$ C.L) and
$B(J/\psi\rightarrow\gamma X)B(X\rightarrow
p\bar{p})=(9.0^{+0.4}_{-1.1}~{\rm (stat.)}^{+1.5}_{-5.0}~{\rm
 (syst.)}\pm2.3~{\rm
 (model)})\times10^{-5}$, respectively. The produce BR for $X(p\bar{p})$ in $\psi^\prime$ decay is first measured to
 be $B(\psi^\prime\rightarrow\gamma X) \times B(X\rightarrow
p\bar{p})=(4.57\pm0.36~{\rm (stat.)}^{+1.23}_{-4.07}~{\rm
(syst.)}\pm1.28~{\rm (model)})\times10^{-6}$   and  the production
ratio of the $X(p\bar{p})$ between $J/\psi$ and $\psi^\prime$
radiative decays is $R=(5.08^{+0.71}_{-0.45}~{\rm
(stat.)}^{+0.67}_{-3.58}~{\rm
 (syst.)}\pm0.12~{\rm
 (model)})\%$.

The mass of the $X(p\bar{p})$ measured in the PWA fit with FSI
effect included is consistent with the $X(1835)$, but the width is
significantly narrower. This indicates that either the $X(p\bar{p})$
and the $X(1835)$ come from different sources, or that interference
effects in the $J/\psi\rightarrow\gamma\pi^+\pi^-\eta^\prime$
process should not be ignored in the determination of the $X(1835)$
mass and width, or that there may be more than one resonance in the
mass peak around 1.83 ${\rm GeV}/c^2$ in
$J/\psi\rightarrow\gamma\pi^+\pi^-\eta^\prime$ decays. When more
$J/\psi$ data are collected at BESIII, more sophisticated analyses,
including a PWA, will be performed for the
$J/\psi\rightarrow\gamma\pi\pi\eta^\prime$ decay channel. A
measurement of the relative production ratios for the $X(1835)$ in
$J/\psi$ and $\psi^\prime$ radiative decays can further clarify on
basis of their production ratios whether or not $X(p\bar{p})$ and
$X(1835)$ are the same states.

 We thank
the accelerator group and computer staff of IHEP for their effort in
producing beams and processing data. We are grateful for support
from our institutes and universities and from these agencies:
Ministry of Science and Technology of China, National Natural
Science Foundation of China, Chinese Academy of Sciences, Istituto
Nazionale di Fisica Nucleare, Russian Foundation for Basic Research,
Russian Academy of Science (Siberian branch), U.S. Department of
Energy,  U.S. National Science Foundation, University of Groningen
(RuG) and the Helmholtzzentrum fuer Schwerionenforschung GmbH (GSI),
and National Research Foundation of Korea.


\ed